&pdflatex

\documentclass[10pt,twocolumn,pra,aps,amssymb,amsmath,tightenlines,showpacs]{revtex4}
\usepackage{graphicx}

\newcommand{\cP}{\ensuremath{\mathcal{P}}}
\newcommand{\cT}{\ensuremath{\mathcal{T}}}
\newcommand{\veps}{\varepsilon}

\begin{document}

\title{Universal spectral behavior of $x^2(ix)^\veps$ potentials}

\author{Carl~M.~Bender\footnote{Current address: Department of Physics, King's
College London, Strand, London, WC2R 2LS, UK\\
{\tt email: cmb@wustl.edu}} and Daniel~W.~Hook\footnote{Current address:
Theoretical Physics, Imperial College, London SW7 2AZ, UK\\
{\tt email: d.hook@imperial.ac.uk}}}

\affiliation{Department of Physics, Washington University, St.~Louis, MO 63130,
USA}

\date{\today}

\begin{abstract}
The $\cP\cT$-symmetric Hamiltonian $H=p^2+x^2(ix)^\veps$ ($\veps$ real) exhibits
a phase transition at $\veps=0$. When $\veps\geq0$, the eigenvalues are all
real, positive, discrete, and grow as $\veps$ increases. However, when $\veps<0$
there are only a finite number of real eigenvalues. As $\veps$ approaches $-1$
from above, the number of real eigenvalues decreases to one, and this eigenvalue
becomes infinite at $\veps=-1$. In this paper it is shown that these qualitative
spectral behaviors are generic and that they are exhibited by the eigenvalues of
the general class of Hamiltonians $H^{(2n)}=p^{2n}+x^2(ix)^\veps$ ($\veps$ real,
$n=1,\,2,\,3,\,\ldots$). The complex classical behaviors of these Hamiltonians
are also examined. 
\end{abstract}

\pacs{11.30.Er, 03.65.Db, 11.10.Ef}

\maketitle

\section{Eigenvalue behaviors}
\label{s1}

\begin{figure}
\begin{center}
\includegraphics[scale=0.40, bb=0 0 583 620]{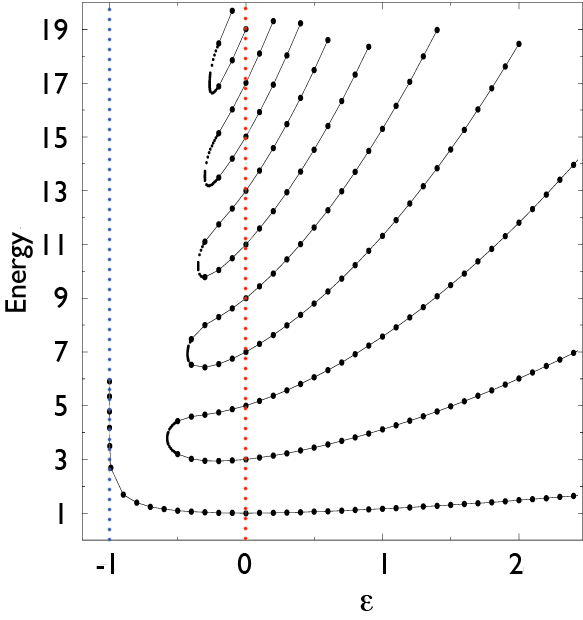}
\end{center}
\caption{Energy levels of the parametric family of Hamiltonians (\ref{e1}). When
$\veps\geq0$, the eigenvalues are all real and positive, and increase with
increasing $\veps$. When $\veps$ decreases below $0$, the eigenvalues disappear
into the complex plane as complex conjugate pairs. Eventually, only one real
eigenvalue remains when $\veps$ is less than about $-0.57$, and as $\veps$
approaches $-1$ from above, this eigenvalue becomes infinite.}
\label{F1}
\end{figure}

The well-studied family of $\cP\cT$-symmetric Hamiltonians \cite{R1,R2,R3,R4,R5}
\begin{equation}
H=p^2+x^2(ix)^\veps\quad(\veps~{\rm real})
\label{e1}
\end{equation}
is a continuous complex deformation in the parameter $\veps$ of the conventional
harmonic-oscillator Hamiltonian $p^2+x^2$. At $\veps=0$ the eigenvalues of $H$
are the usual harmonic-oscillator eigenvalues $E_n=2n+1$ ($n=0,\,1,\,2,\,
\dots$), but as $\veps$ varies away from $0$, the eigenvalues exhibit several
characteristic qualitative behaviors: When $\veps$ is positive, the eigenvalues
are all real, positive, and discrete, and the eigenvalues increase as $\veps$
increases. When $\veps$ is negative, most of the eigenvalues are complex. As
$\veps$ decreases from 0 to $-1$, the eigenvalues sequentially become pairwise
degenerate and enter the complex plane as complex-conjugate pairs. As $\veps$
continues to decrease, eventually only one real eigenvalue remains, and this
eigenvalue becomes infinite at $\veps=-1$ \cite{R6}. This behavior is displayed
in Fig.~\ref{F1}.

The parametric region $\veps\geq0$ is called the {\it region of unbroken $\cP
\cT$ symmetry} because all of the eigenfunctions of the Hamiltonian are
simultaneously eigenstates of $\cP\cT$; the parametric region $\veps<0$ is
called the {\it region of broken $\cP\cT$ symmetry} because the eigenfunctions
of the Hamiltonian are not all eigenstates of $\cP\cT$ \cite{R7}. A phase
transition at $\veps=0$ separates the regions of unbroken and broken symmetry.
Parametric families of $\cP\cT$-symmetric Hamiltonians typically exhibit a phase
transition with entirely real eigenvalues on one side and complex eigenvalues on
the other. The $\cP\cT$ phase transition has been observed repeatedly in
laboratory experiments
\cite{R8,R9,R10,R11,R12,R13,R14,R15,R16,R17,R18,R19,R20,R21,R22}.

The purpose of this paper is to report our observation that the qualitative
features of the eigenvalues of $H$ in (\ref{e1}), as shown in Fig.~\ref{F1},
are generic; that is, the same qualitative behavior is displayed by the general
family of Hamiltonians
\begin{equation}
H^{(2n)}=p^{2n}+x^2(ix)^\veps\quad(\veps~{\rm real};\,n=1,\,2,\,3,\,\ldots)
\label{e2}
\end{equation}
for which $H$ in (\ref{e1}) is just $H^{(2)}$. For example, when $n=2$, the
time-independent eigenvalue problem $H^{(4)}\psi=E\psi$ in coordinate space
becomes the {\it fourth-order} differential equation
\begin{equation}
\frac{d^4}{dx^4}\psi(x)+x^2(ix)^\veps\psi(x)=E\psi(x).
\label{e3}
\end{equation}
To determine the boundary conditions on $\psi(x)$, we begin with the requirement
that $\psi(x)$ vanish exponentially fast as $|x|\to\infty$ on the real-$x$ axis
when $\veps=0$, so the eigenvalues are the same as those of the conventionally
Hermitian quartic-anharmonic-oscillator Hamiltonian $p^2+x^4$ \cite{R23}. Then,
this eigenvalue problem is extended into the complex domain by allowing the
parameter $\veps$ to range continuously from $-1$ to $+\infty$. As we explain in
Sec.~\ref{s2}, this analysis determines the boundary conditions on $\psi(x)$ in
the complex-$x$ plane for real $-1<\veps<\infty$.

\begin{figure}[t!]
\begin{center}
\includegraphics[scale=0.23, bb=0 0 1000 995]{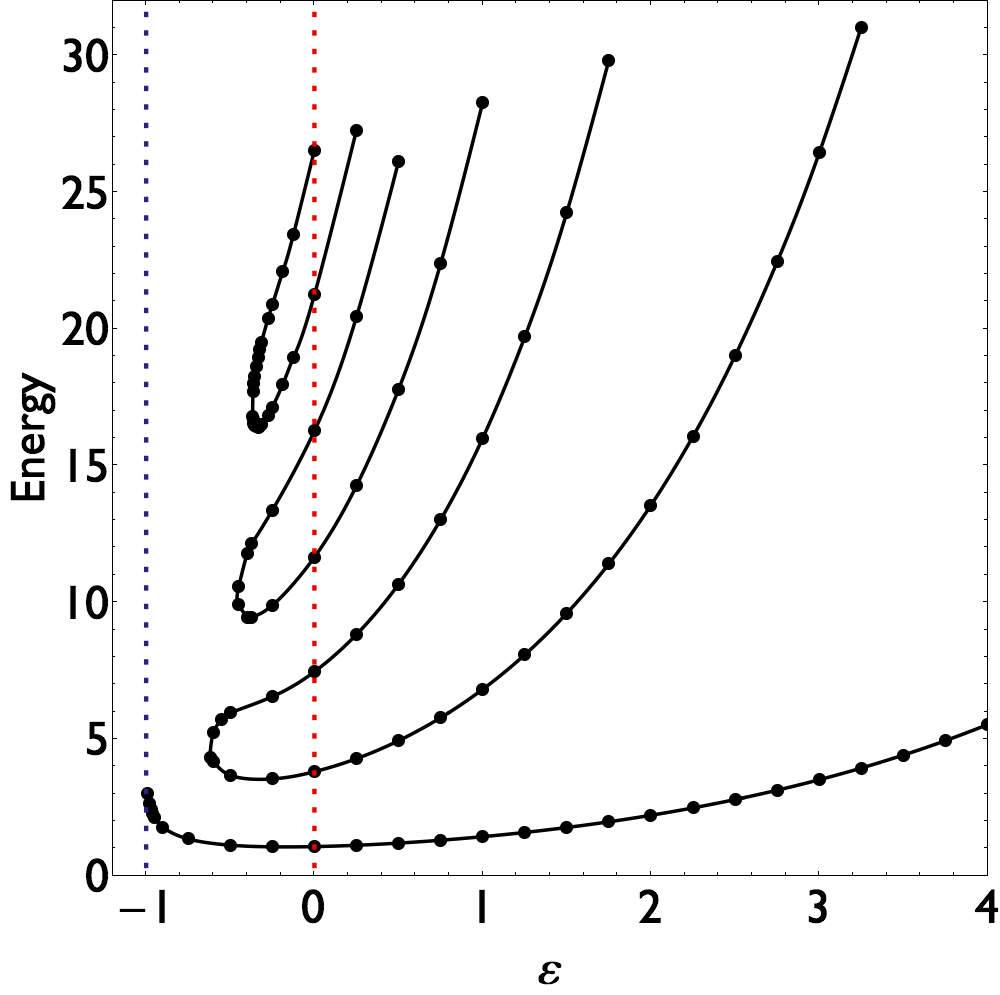}
\end{center}
\caption{Eigenvalues of the $\cP\cT$-symmetric Hamiltonian $H^{(4)}$ in
(\ref{e2}) plotted as functions of $\veps$ for $-1<\veps<4$. The eigenvalue
behavior agrees qualitatively with that in Fig.~\ref{F1} for the case of
$H^{(2)}$.}
\label{F2}
\end{figure}

The eigenvalues of $H^{(4)}$ are plotted in Fig.~\ref{F2}. The behaviors of the
eigenvalues in Fig.~\ref{F2} are qualitatively similar to the eigenvalues in
Fig.~\ref{F1}. The eigenvalues in both cases are real, positive, and discrete
for $\veps\geq0$ and they become complex below the phase transition at $\veps=
0$; at $\veps=-1$ there are no eigenvalues at all.

\begin{figure}
\begin{center}
\includegraphics[scale=0.23, bb=0 0 1000 1005]{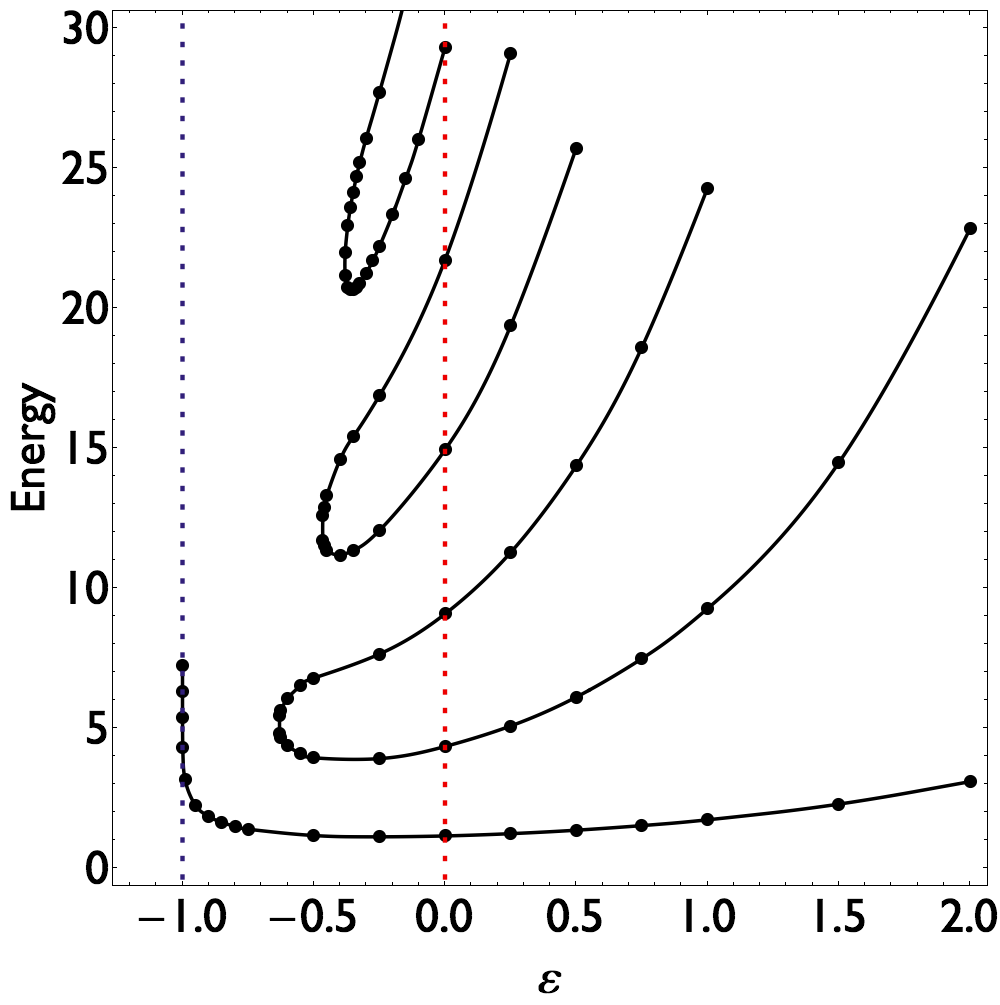}
\end{center}
\caption{Eigenvalues of the $\cP\cT$-symmetric Hamiltonian $H^{(6)}$ in
(\ref{e2}) plotted as functions of $\veps$ for $-1<\veps<2$. The eigenvalue
behavior is qualitatively similar to that in Figs.~\ref{F1} and \ref{F2} for
$H^{(2)}$ and $H^{(4)}$.}
\label{F3}
\end{figure}

When $n=3$, the time-independent eigenvalue problem $H^{(6)}\psi=E\psi$ is the
{\it sixth-order} differential equation
\begin{equation}
-\frac{d^6}{dx^6}\psi(x)+x^2(ix)^\veps\psi(x)=E\psi(x).
\label{e4}
\end{equation}
When $\veps=0$, the eigenvalues are the same as those of the conventionally
Hermitian sextic-anharmonic-oscillator Hamiltonian $p^2+x^6$. Observe that for
$-1<\veps<\infty$ the eigenvalues, which are shown in Fig.~\ref{F3}, behave
qualitatively like those in Figs.~\ref{F1} and \ref{F2}.

This paper is organized very simply: In Sec.~\ref{s2} we explain the asymptotic
and numerical analysis that leads to the results presented in Figs.~\ref{F2} and
\ref{F3}. In Sec.~\ref{s3} we study the complex classical trajectories
associated with the Hamiltonians $H^{(2)}$, $H^{(4)}$, and $H^{(6)}$ for a
variety of values of $\veps$. These trajectories demonstrate some qualitative
differences between these Hamiltonians. Finally, in Sec.~\ref{s4} we make some
brief concluding remarks.

\section{Asymptotic and Numerical Analysis}
\label{s2}

In this section we explain the techniques for calculating the eigenvalues that
are graphed in Figs.~\ref{F1}--\ref{F3}. We begin by reviewing the procedure for
finding the eigenvalues of $H^{(2)}$.

\subsection{Review of the case $H^{(2)}$}
\label{s21}

The properties of the eigenvalues of $H^{(2)}$ were first elucidated in
Ref.~\cite{R1}. For completeness we give a brief review of the techniques that
were used to construct an accurate plot of the eigenvalues.

The eigenvalue equation (the time-independent Schr\"odinger equation) for
$H^{(2)}$ reads
\begin{equation}
-\psi''(x)+x^2(ix)^\veps\psi(x)=E\psi(x).
\label{e5}
\end{equation}
This differential equation has many different sets of eigenvalues, which are
distinguished by the boundary conditions imposed on the eigenfunctions. Thus, it
is crucial to identify the specific boundary conditions on the eigenfunctions
$\psi(x)$ that give rise to the eigenvalues plotted in Fig.~\ref{F1}.

WKB theory can be used to obtain the possible asymptotic behaviors of $\psi(x)$
for large $|x|$:
\begin{equation}
\psi(x)\sim\exp\left(\pm x^{2+\veps/2}e^{i\pi\veps/4}\right)\quad(|x|\to\infty),
\label{e6}
\end{equation}
where we have used $i=e^{i\pi/2}$. Since (\ref{e5}) is a second-order
differential equation, there are two solutions; we want the solution that decays
exponentially as $|x|\to\infty$ on the real-$x$ axis when $\veps=0$.

To understand what happens when $\veps\neq0$, we let $x=re^{i\theta}$ and
require that as $r\to\infty$,
\begin{equation}
\psi(r)\sim\exp\left\{-\frac{2}{4+\veps} r^{2+\veps/2}e^{i[\theta(2+\veps/2)+\pi
\veps/4]}\right\}.
\label{e7}
\end{equation}
Note that $\psi(r)$ vanishes in a Stokes wedge. At the center of the this wedge
the phase, which is the expression in square brackets in (\ref{e7}), vanishes.
This gives the angular location of the center of the wedge:
\begin{equation}
\theta_R=-\pi\veps/(8+2\veps).
\label{e8}
\end{equation}
We use the notation $\theta_R$ here to indicate that this is the {\it right}
Stokes wedge. There is also a {\it left} Stokes wedge, which is the mirror image
of the right Stokes wedge; the angle of the center of the left Stokes wedge is
called $\theta_L$. The mirror for this problem is the imaginary axis; the
imaginary axis serves as the mirror because the Hamiltonian $H^{(2)}$ is $\cP
\cT$ symmetric. Note that $\theta_R=0$ at $\veps=0$ and that $\theta_R\to-\pi/2$
as $\veps\to\infty$.

The opening angle $\Delta\theta$ of the right Stokes wedge (and of the left
Stokes wedge) is determined by requiring that the phase in square brackets in
(\ref{e7}) is $\pm\pi/2$. This gives
\begin{equation}
\Delta\theta=2\pi/(4+\veps).
\label{e9}
\end{equation}
The opening angles of the Stokes wedges shrink to $0$ as $\veps\to\infty$.
However, as $\veps\to-1$, the opening angles expand to $\Delta\theta/2=\pi/3$.
Thus, since $\theta_R=\pi/6$ at $\veps=-1$, we see that the upper edge of the
right wedge is at $\theta_R+\Delta\theta/2=\pi/2$. At this value of $\veps$ the
left and right wedges meet and the eigenvalue problem disappears because the
integration contour can be deformed to infinity. This is an intuitive way to
understand the rigorous result of Herbst that the spectrum becomes null when
$\veps=-1$ \cite{R6}.

The numerical calculation of the eigenvalues uses the standard {\it shooting
method}. We integrate numerically with respect to the $r$ variable down the
centers of the left and right wedges and join the two solutions at the origin.
The differential equation in the right wedge in the $r$ variable reads
\begin{equation}
\psi_R''(r)=\left(r^{2+\veps}+Ee^{2i\theta_R}\right)\psi_R(r)=Q(r)\psi_R(r).
\label{e10}
\end{equation}
The numerical integration begins at a large value of $r$, say $r=r_0$. To
determine the appropriate initial conditions at $r=r_0$, we use the WKB
approximation for large $r_0$:
\begin{equation}
\psi_{\rm WKB}\left(r_0\right)\sim KQ^{-1/4}\left(r_0\right)\exp\left[-
\int_0^{r_0}dr\sqrt{Q(r)}\right],
\label{e11}
\end{equation}
where $K$ is an arbitrary constant. We then adjust the constant $K$ so that 
\begin{equation}
\psi_R\left(r_0\right)=1~{\rm and}~\psi_R'\left(r_0\right)=-\sqrt{Q\left(r_0
\right)}.
\label{e12}
\end{equation}
For the eigenvalues shown in Fig.~\ref{F1} we choose $r_0=12$.

We patch the left and right solutions at the origin as follows: In the left
wedge the eigenfunction that vanishes exponentially for large argument $s$ is
$\alpha\psi_L(s)$, and in the right wedge the eigenfunction that vanishes
exponentially for large $r$ is $\beta\psi_R(r)$, where $\alpha$ and $\beta$
are arbitrary constants. Thus, the patching conditions are
\begin{equation}
\alpha\psi_L(0)=\beta\psi_R(0),~~\alpha e^{-i\theta_L}\psi_L'(0)=\beta
e^{-i\theta_R}\psi_R'(0).
\label{e13}
\end{equation}
These are homogeneous linear equations, and thus by Cramer's rule the
determinant $D$ must vanish, where
\begin{equation}
D={\rm det}\left(\begin{array}{cc} \psi_R(0) & -\psi_L(0) \\
e^{-i\theta_R}\psi_R'(0) & -e^{-i\theta_L}\psi_L(0) \\ \end{array}\right).
\label{e14}
\end{equation}

The shooting procedure consists of adjusting the eigenvalue $E$ so as to
minimize $|D|$. This numerical procedure is highly efficient and gives a rapidly
convergent sequence of numerical approximations to $E$.

\subsection{Case $H^{(4)}$}
\label{s22}

Let us consider the fourth-order differential-equation eigenvalue problem in
(\ref{e3}). A local analysis of this differential equation for large $|x|$ shows
that, apart from multiplicative algebraic corrections, the four possible
asymptotic behaviors of $\psi(x)$ are
\begin{equation}
\psi(x)\sim\exp\left(\frac{4\omega}{6+\veps}x^{(6+\veps)/4}e^{i\pi\veps/8}
\right)\quad(|x|\to\infty),
\label{e15}
\end{equation}
where $\omega^4=-1$. We must exclude two of these four solutions because they 
grow exponentially when $|x|\gg1$.

We now describe the Stokes wedges that are associated with the solutions. We let
$x=re^{i\theta}$, and in the right Stokes wedge we set $\theta=\theta_R$.
Neglecting for the moment the effect of $\omega$, the phase angle in (\ref{e15})
is $(6+\veps)\theta_R/4+\pi\veps/8$. This angle vanishes at the center of the
Stokes wedge. Thus,
\begin{equation}
\theta_R=-\pi\veps/(12+2\veps).
\label{e16}
\end{equation}
As in the case of $H^{(2)}$, the center of the wedge is located at $\theta_R=0$ 
when $\veps=0$, and the center rotates downward (clockwise) to $-\pi/2$ as
$\veps\to\infty$.

As before, the opening angle $\Delta\theta$ of the Stokes wedge is determined by
the condition that the phase has the value $\pm\pi/2$. This gives
\begin{equation}
\Delta\theta=4\pi/(6+\veps).
\label{e17}
\end{equation}
Thus, the opening angle is $2\pi/3$ at $\veps=0$ and the opening angle vanishes
as $\veps\to\infty$.

We now observe one of the universal features of the Hamiltonians in (\ref{e2}).
At $\veps=-1$, $\theta_R=\pi/10$ and $\Delta\theta/2=4\pi/10$. Thus, the upper
edge of the right wedge is at $\theta_R+\Delta\theta/2=\pi/2$. By $\cP\cT$
symmetry, there is a left wedge (a mirror reflection of the right wedge) in the
left-half plane. The center of the left wedge is at the angle $\theta_L=-\pi+
\theta_R$. Hence, when $\veps=-1$, the left and the right wedges touch, the
eigenvalue problem collapses, and the spectrum becomes null. (This is the
generalization of the effect that Herbst found for the Hamiltonian $H=p^2\pm ix$
\cite{R6}.) This collapse of the eigenvalue problem is a universal phenomenon
and always occurs at $\veps=-1$ for all Hamiltonians in (\ref{e2}).

We now explain the numerical procedure for finding the eigenvalues in
Fig.~\ref{F2}. In the right wedge we substitute $x=re^{i\theta_R}$ in the
differential equation (\ref{e3}) and obtain
\begin{equation}
\psi_R^{(4)}(r)=-\left(r^{2+\veps}+Ee^{4i\theta_R}\right)\psi_R(r)=-Q(r)\psi_R
(r).
\label{e18}
\end{equation}
There are two solutions that grow exponentially with $r$ and two solutions that
vanish exponentially with $r$.

Until now, $\omega$ has not played any part in the analysis. However, it plays a
crucial role in determining the boundary conditions. We integrate numerically
downward in $r$ from a large value of $r=r_0$. To fix the initial conditions we
use the WKB approximation for fourth-order differential equations \cite{R24}:
\begin{equation}
\psi_{\rm WKB}\left(r_0\right)\sim KQ^{-\frac{3}{8}}\left(r_0\right)\exp\left[
\omega\int_0^{r_0}dr\,Q^{1/4}(r)\right],
\label{e19}
\end{equation}
where $\omega^4=-1$. We choose the constant $K$ so that
\begin{equation}
\psi_R\left(r_0\right)=1.
\label{e20}
\end{equation}
Then, using $\omega_1=e^{3\pi i/4}$ and $\omega_2=e^{-3\pi i/4}$ so that the
solution decays exponentially, we take
\begin{equation}
\psi_{R,j}^{(k)}\left(r_0\right)=\omega_j^k Q^{(k/4)}\left(r_0\right),
\label{e21}
\end{equation}
where $\psi_{R,j}^{(k)}$ is the $k$th derivative with respect to $r$. We then
choose a large number for $r_0$, say $r_0=12$. The solution in the left wedge is
\begin{equation}
\psi_L=\alpha_1\psi_{L,1}+\alpha_2\psi_{L,2}
\label{e22}
\end{equation}
and the solution in the right wedge is
\begin{equation}
\psi_R=\beta_1\psi_{R,1}+\beta_2\psi_{R,2}.
\label{e23}
\end{equation}

\begin{widetext}
We must now patch the left and right solutions at $r=0$. To do so we require
that the eigenfunctions $\psi_L$ and $\psi_R$ be continuous and have continuous
first, second, and third derivatives at $r=0$. These four conditions are written
as a system of four linear equations for $\alpha_1$, $\alpha_2$, $\beta_1$, and
$\beta_2$:
\begin{eqnarray}
\alpha_1\psi_{L,1}(0)+\alpha_2\psi_{L,2}(0) &=&
\beta_1\psi_{R,1}(0) +\beta_2 \psi_{R,2}(0),\nonumber\\
\alpha_1e^{-i\theta_L}\psi_{L,1}'(0)+\alpha_2 e^{-i\theta_L}\psi_{L,2}'(0) &=&
\beta_1 e^{-i\theta_R}\psi_{R,1}'(0)+\beta_2 e^{-i\theta_R}\psi_{R,2}'(0),
\nonumber\\
\alpha_1e^{-2i\theta_L}\psi_{L,1}''(0)+\alpha_2e^{-2i\theta_L}\psi_{L,2}''(0)&=&
\beta_1e^{-2i\theta_R}\psi_{R,1}''(0)+\beta_2e^{-2i\theta_R}\psi_{R,2}''(0),
\nonumber\\
\alpha_1e^{-3i\theta_L}\psi_{L,1}'''(0)+\alpha_2 e^{-3i\theta_L}\psi_{L,2}'''(0)
&=& \beta_1e^{-3i\theta_R}\psi_{R,1}'''(0)+\beta_2 e^{-3i\theta_R}\psi_{R,2}'''
(0).
\label{e24}
\end{eqnarray}
Because these patching conditions are homogeneous equations, they translate via
Cramer's rule to the condition that the following $4\times4$ determinant of the
coefficients vanishes:
\begin{equation}
D={\rm det}\left(\begin{array}{cccc}
\psi_{R,1}(0) & \psi_{R,2}(0) & -\psi_{L,1}(0) & -\psi_{L,2}(0) \\
e^{-i\theta_R}\psi_{R,1}'(0) & e^{-i\theta_R}\psi_{R,2}'(0) &
-e^{-i\theta_L}\psi_{L,1}'(0) & -e^{-i\theta_L}\psi_{L,2}'(0) \\
e^{-2i\theta_R}\psi_{R,1}''(0) & e^{-2i\theta_R}\psi_{R,2}''(0) &
-e^{-2i\theta_L}\psi_{L,1}''(0) & -e^{-2i\theta_L}\psi_{L,2}''(0) \\
e^{-3i\theta_R}\psi_{R,1}'''(0) & e^{-3i\theta_R}\psi_{R,2}'''(0) &
-e^{-3i\theta_L}\psi_{L,1}'''(0) & -e^{-3i\theta_L}\psi_{L,2}'''(0) \\
\end{array}\right).
\label{e25}
\end{equation}
\end{widetext}
The objective of the numerical shooting procedure is to find the values of
$E$ that minimize this determinant.

\subsection{Case $H^{(6)}$}
\label{s23}

For the eigenvalue problem in (\ref{e4}) the possible asymptotic behaviors for
large $|x|$ (excluding algebraic corrections) are given by
\begin{equation}
\psi(x)\sim\exp\left(\frac{6\omega}{8+\veps}x^{(8+\veps)/6}e^{i\pi\veps/12}
\right)\quad(|x|\to\infty),
\label{e26}
\end{equation}
where $\omega^6=1$.

There are six solutions, but we seek the three solutions that vanish as $|x|\to
\infty$. We substitute $x=re^{i\theta_R}$ in (\ref{e26}), and we require that
the solution vanish as $r\to\infty$. Apart from $\omega$, the phase angle in the
asymptotic behavior vanishes when 
\begin{equation}
\theta_R=-\pi\veps/(16+2\veps).
\label{e27}
\end{equation}
This determines the center of the right wedge. At $\veps=0$, $\theta_R=0$ and
as $\veps\to\infty$, $\theta_R\to-\pi/2$, just as before.

The opening angle $\Delta\theta$ of the wedge is determined by requiring that
the phase have the value $\pm\pi/2$. This gives
\begin{equation}
\Delta\theta=6\pi/(8+\veps).
\label{e28}
\end{equation}
At $\veps=0$ the opening angle is $3\pi/4$ and this angle vanishes as $\veps\to
\infty$.

From (\ref{e27}) and (\ref{e28}) we see that at $\veps=-1$, $\theta_R=\pi/14$
and $\Delta\theta/2=6\pi/14$. Thus, the upper edge of the wedge is at $\theta_R+
\Delta\theta/2=\pi/2$. At this value of $\veps$, the left and right wedges touch
and the eigenvalue problem disappears, just as in the case of $H^{(2)}$ and
$H^{(4)}$.

To find the eigenvalues using the numerical shooting method, we integrate down
the center of the left and right wedges and patch the solutions at the origin.
In the right wedge the differential equation in the $r$ variable is
\begin{equation}
\psi_R^{(6)}(r)=\left(r^{2+\veps}+Ee^{6i\theta_R}\right)\psi_R(r)=Q(r)\psi_R(r).
\label{e29}
\end{equation}
In each wedge there are three exponentially decaying solutions and three
exponentially growing solutions. We select the decaying solutions by applying
the appropriate boundary conditions at a large value of $r=r_0=12$. To identify
the appropriate initial conditions for the numerical integration, we use the WKB
approximation for large $r_0$:
\begin{equation}
\psi_{\rm WKB}\left(r_0\right)\sim KQ^{-\frac{5}{12}}\left(r_0\right)\exp\left[
\omega\int_0^{r_0}dr\,Q^{1/6}(r)\right],
\label{e30}
\end{equation}
where $\omega_1=-1$, $\omega_2=e^{2\pi i/3}$, and $\omega_3=e^{-2\pi i/3}$. We
adjust the constant $K$ such that
\begin{equation}
\psi_R\left(r_0\right)=1.
\label{e31}
\end{equation}
Then, 
\begin{equation}
\psi_{R,j}^{(k)}\left(r_0\right)=\omega_j^k Q^{(k/6)}\left(r_0\right).
\label{e32}
\end{equation}
The solution in the left wedge is 
\begin{equation}
\psi_L=\alpha_1\psi_{L,1}+\alpha_2\psi_{L,2}+\alpha_3\psi_{L,3}
\label{e33}
\end{equation}
and the solution in the right wedge is
\begin{equation}
\psi_R=\beta_1\psi_{R,1}+\beta_2\psi_{R,2}+\beta_3\psi_{R,3}.
\label{e34}
\end{equation}

The patching conditions give a vanishing determinant via Cramer's rule, where
the determinant is
\begin{widetext}
\begin{equation}
D={\rm det}\left(\begin{array}{cccccc}
\psi_{R,1}(0) & \psi_{R,2}(0) & \psi_{R,3}(0) &
-\psi_{L,1}(0) & -\psi_{L,2}(0) & -\psi_{L,3}(0) \\
e^{-i\theta_R}\psi_{R,1}'(0) & e^{-i\theta_R}\psi_{R,2}'(0) &
e^{-i\theta_R}\psi_{R,3}'(0) & -e^{-i\theta_L}\psi_{L,1}'(0) &
-e^{-i\theta_L}\psi_{L,2}'(0) & -e^{-i\theta_L}\psi_{L,3}'(0) \\
e^{-2i\theta_R}\psi_{R,1}''(0) & e^{-2i\theta_R}\psi_{R,2}''(0) &
e^{-2i\theta_R}\psi_{R,3}''(0) & -e^{-2i\theta_L}\psi_{L,1}''(0) &
-e^{-2i\theta_L}\psi_{L,2}''(0) & -e^{-2i\theta_L}\psi_{L,3}''(0) \\
e^{-3i\theta_R}\psi_{R,1}'''(0) & e^{-3i\theta_R}\psi_{R,2}'''(0) &
e^{-3i\theta_R}\psi_{R,3}'''(0) & -e^{-3i\theta_L}\psi_{L,1}'''(0) &
-e^{-3i\theta_L}\psi_{L,2}'''(0) & -e^{-3i\theta_L}\psi_{L,3}'''(0) \\
e^{-4i\theta_R}\psi_{R,1}^{(4)}(0) & e^{-4i\theta_R}\psi_{R,2}^{(4)}(0) &
e^{-4i\theta_R}\psi_{R,3}^{(4)}(0) & -e^{-4i\theta_L}\psi_{L,1}^{(4)}(0) &
-e^{-4i\theta_L}\psi_{L,2}^{(4)}(0) & -e^{-4i\theta_L}\psi_{L,3}^{(4)}(0) \\
e^{-5i\theta_R}\psi_{R,1}^{(5)}(0) & e^{-5i\theta_R}\psi_{R,2}^{(5)}(0) &
e^{-5i\theta_R}\psi_{R,3}^{(5)}(0) & -e^{-5i\theta_L}\psi_{L,1}^{(5)}(0) &
-e^{-5i\theta_L}\psi_{L,2}^{(5)}(0) & -e^{-5i\theta_L}\psi_{L,3}^{(5)}(0) \\
\end{array}\right).
\label{e35}
\end{equation}
\end{widetext}
We shoot so as to minimize this determinant.

\section{Classical Trajectories}
\label{s3}

\begin{figure}
\begin{center}
\includegraphics[scale=0.23, bb=0 0 1000 2051]{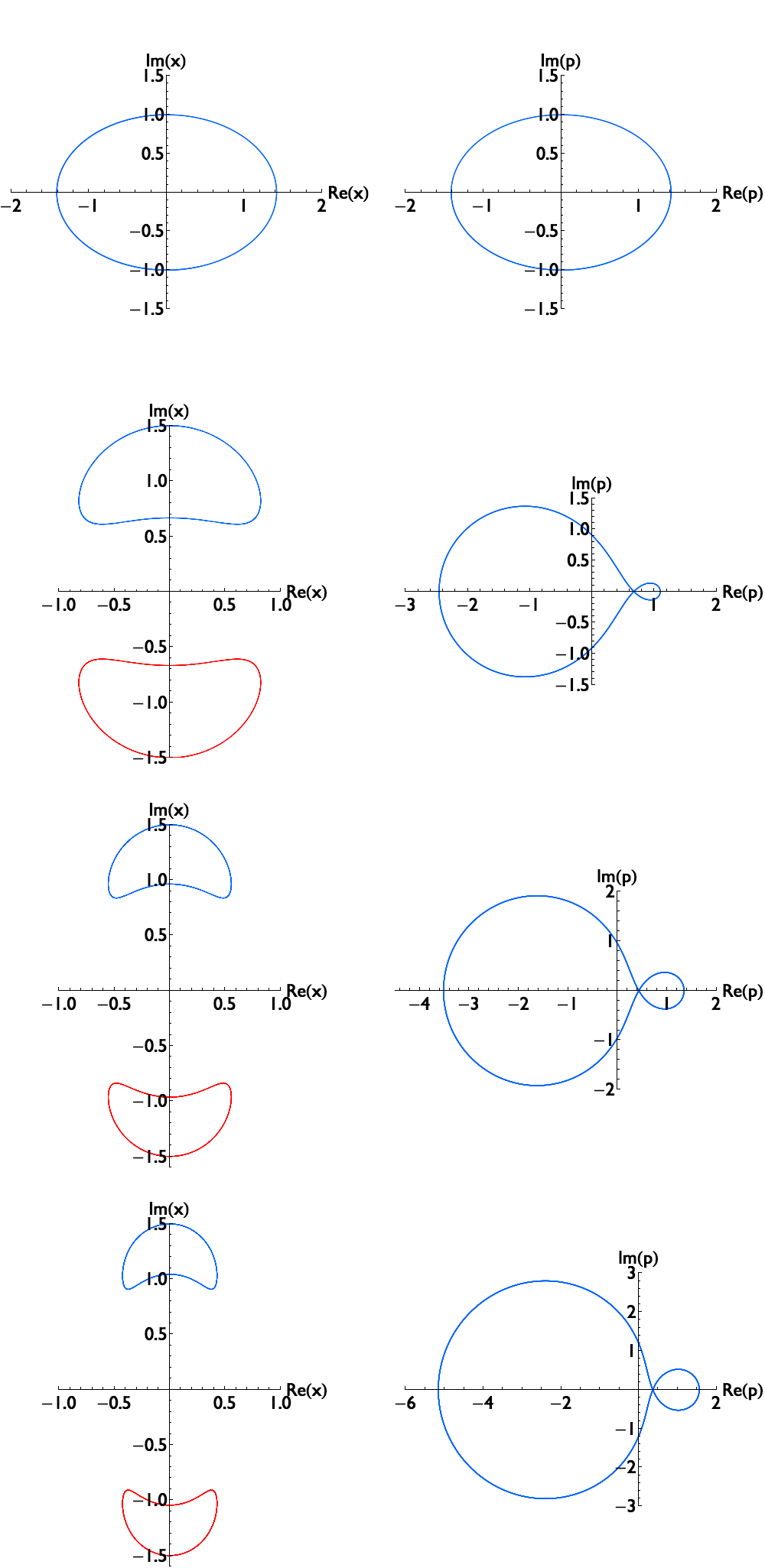}
\end{center}
\caption{Complex classical trajectories of the Hamiltonians $H^{(2)}=p^2+x^2(ix
)^\veps$, where $\veps=0$ (first row), $\veps=2$ (second row), $\veps=4$ (third
row), and $\veps=6$ (fourth row). The left column is the projection of the
classical trajectory in the complex-$x$ plane and the right column is the
projection of the trajectory in the complex-$p$ plane. In every case the
classical energy is 1. For the cases $\veps=2$, $4$, and $6$ {\it two}
trajectories are shown in the left column (blue and red online). The
corresponding complex-$p$-plane trajectories do not cross, but rather lie on a
two-sheeted Riemann surface. The $p$-plane trajectories appear to be coincident
because they are projected onto a single complex plane.}
\label{F4}
\end{figure}

\begin{figure}
\begin{center}
\includegraphics[scale=0.23, bb=0 0 1000 1884]{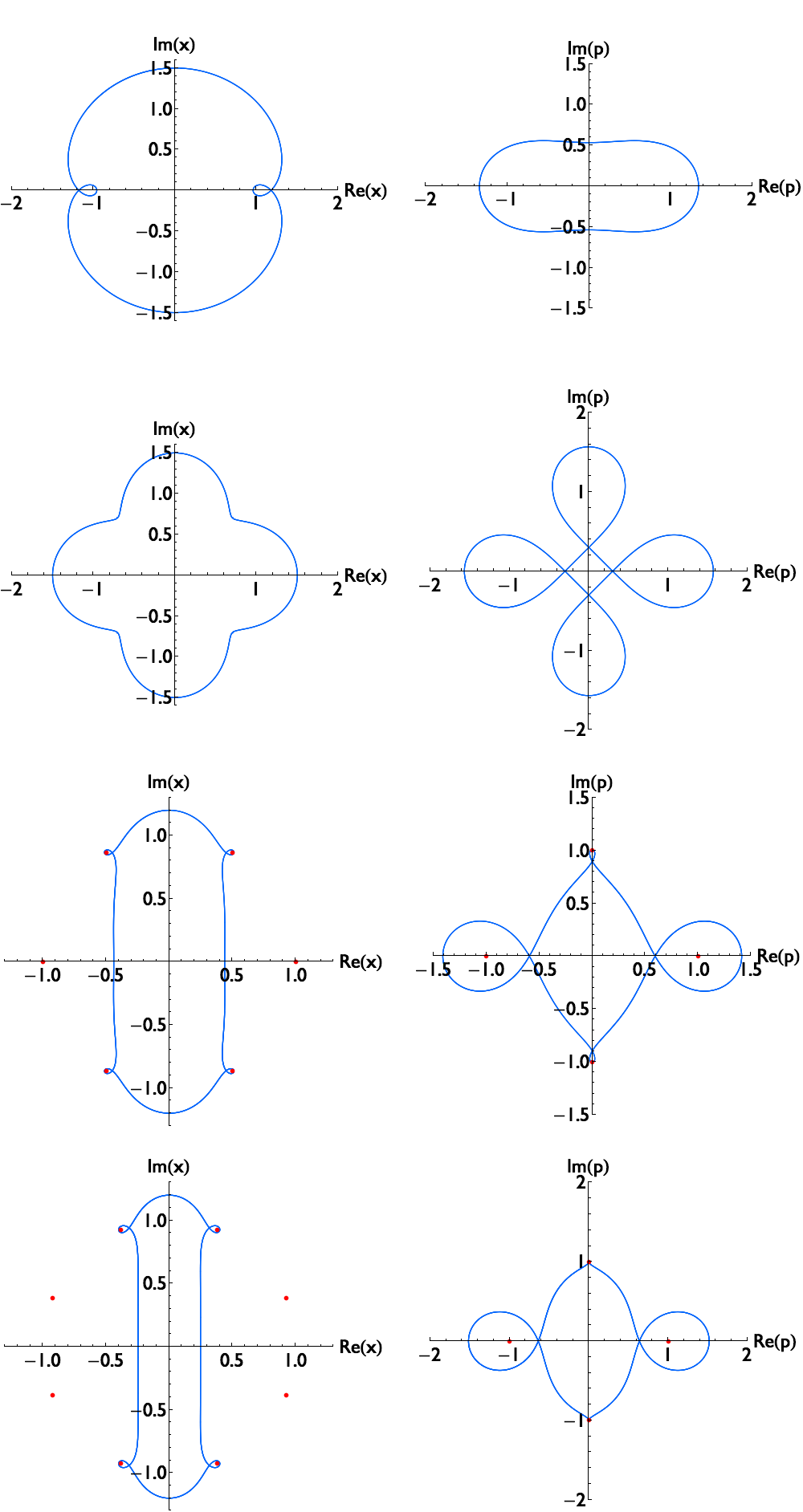}
\end{center}
\caption{Same as Fig.~\ref{F4} panel except that $n=2$. However, in this case we
do not observe up-down trajectory pairing in the complex-$x$ plane.}
\label{F5}
\end{figure}

\begin{figure}
\begin{center}
\includegraphics[scale=0.22, bb=0 0 1000 2051]{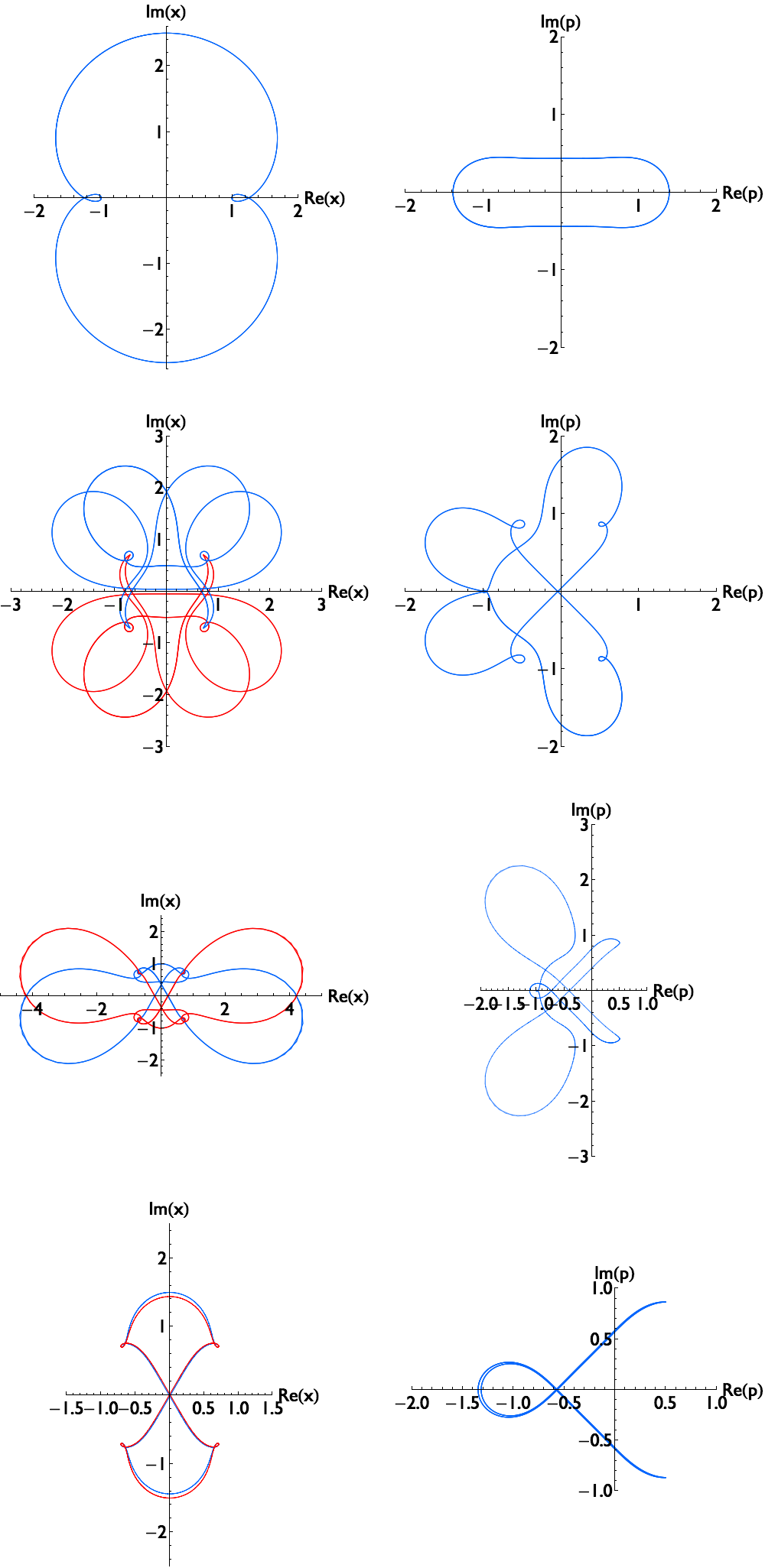}
\end{center}
\caption{Complex classical trajectories of the Hamiltonians $H^{(6)}=p^6+x^2(ix
)^\veps$, where $\veps=0$ (first row), $\veps=2$ (second, third, and fourth
rows). The left column is the projection of the classical trajectory in the
complex-$x$ plane and the right column is the projection of the trajectory in
the complex-$p$ plane. In every case the classical energy is 1. In rows 2, 3,
and 4 three distinct topologies are shown. As in Fig.~\ref{F4}, these
trajectories come in up-down symmetric pairs (blue and red online). The
corresponding complex-$p$-plane trajectories do not cross, but rather lie on a
six-sheeted Riemann surface.}
\label{F6}
\end{figure}

\begin{figure}
\begin{center}
\includegraphics[scale=0.22, bb=0 0 1000 1524]{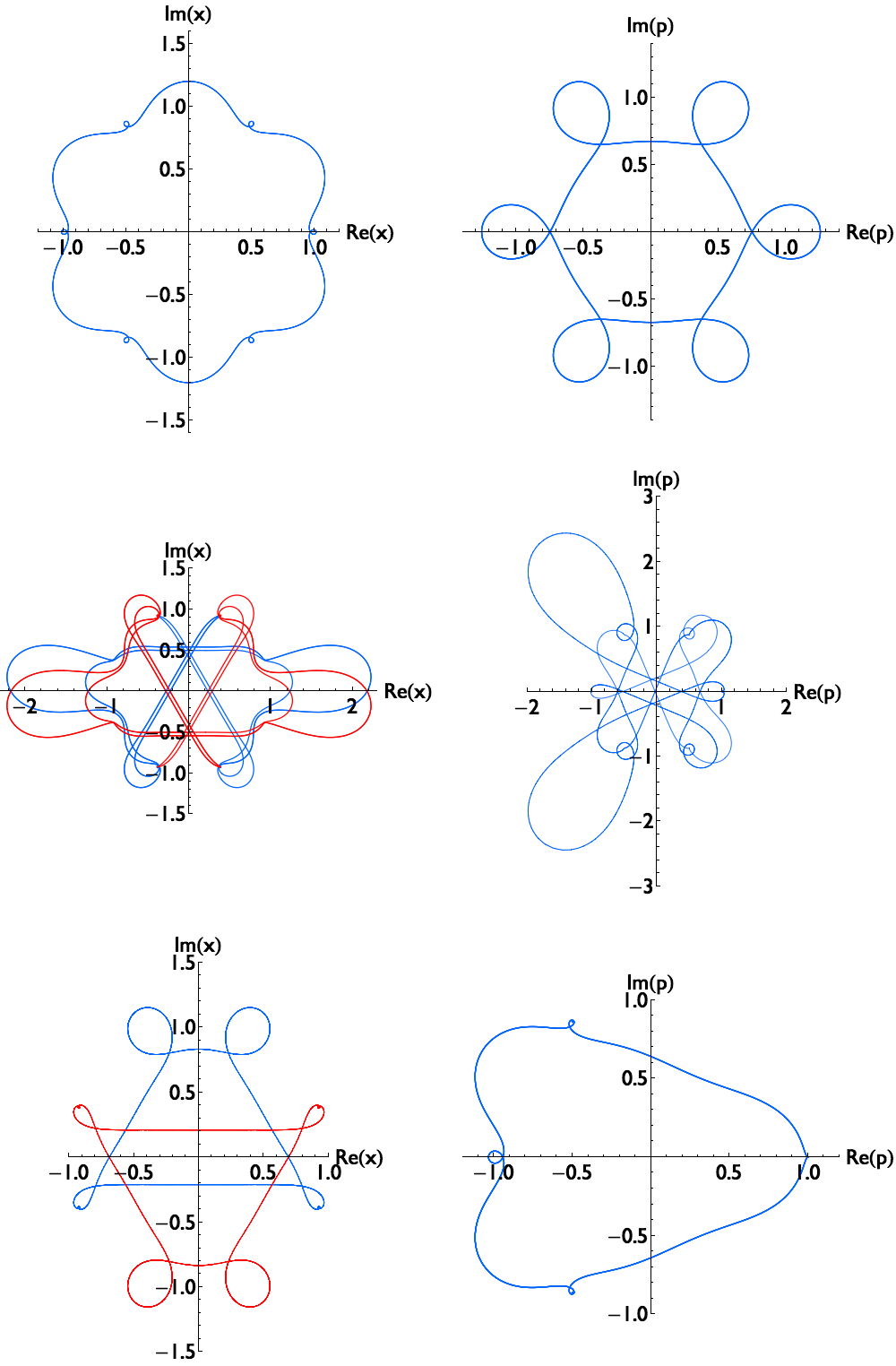}
\end{center}
\caption{Same as Fig.~\ref{F6} except that $\veps=4$ (first row), and $\veps=6$ 
(second and third rows). Two distinct topologies are shown for $\veps=6$.}
\label{F7}
\end{figure}

While the eigenvalues of the quantum-mechanical Hamiltonians $H^{(2n)}$ are
qualitatively similar, the classical-mechanical trajectories for these
Hamiltonians exhibit marked differences. Using the numerical techniques 
described in Ref.~\cite{R2}, we have studied the complex classical trajectories
associated with the Hamiltonians in (\ref{e2}) for $\veps=0$, $2$, $4$, and $6$.
In Fig.~\ref{F4} we plot four trajectories in $x$ space and in $p$ space for
$n=1$ for these values of $\veps$ and in Fig.~\ref{F5} we show corresponding
plots for $n=2$. Figures \ref{F6} and \ref{F7} display the corresponding plots
for $n=3$, and in these figures we can see the emergence of new kinds of
topological structures. Figures \ref{F4}-\ref{F7} show a progressively
increasing topological complexity, which arises because of the higher powers of
$p$ in the Hamiltonians. It is interesting that the pattern of the eigenvalues
in Figs.~\ref{F1}-\ref{F3} shows no corresponding qualitative changes.

\section{Concluding Remarks}
\label{s4}

We have shown that the key qualitative features of the eigenvalue problem
associated with $H^{(2n)}$ in (\ref{e2}) are universal. In particular, (i) the
critical point between the regions of broken and unbroken symmetry is always
at $\veps=0$, (ii) the lower end of the broken region is always at $\veps=-1$ at
which point the spectrum becomes null, and (iii) the eigenvalues are positive
and rise with increasing $\veps$ in the unbroken region.

Qualitative differences between these Hamiltonians are much easier to spot at
the classical level. We see that as $n$ increases, the classical trajectories 
develop new and richer topologies. Near the turning points the trajectories
share the same features as those studied in Ref.~\cite{R25}, in which the
classical paths for higher-derivative Hamiltonians were also investigated.

CMB thanks the Leverhulme Foundation and the U.S.~Department of Energy for
financial support. Mathematica was used to perform numerical calculations.

\end{document}